\documentclass[aps,twocolumn,superscriptaddress,prl]{revtex4-2}

\usepackage{bm,graphicx,textcomp,amssymb,amsmath,dcolumn,color}

\usepackage{multirow}
\usepackage{tabularx}
\usepackage{ctable}
   
\usepackage[utf8]{inputenc}
\usepackage[T1]{fontenc}

\usepackage{bbm}
\usepackage{bbold}   
\usepackage{mathtools}

\usepackage{ulem}
\usepackage{comment}

\usepackage{lipsum}  
\usepackage{graphicx,stackengine}
\newcommand\underlay[4]{%
  \stackengine{0pt}%
  {\kern#2\includegraphics[height=#1]{#4}}%
  {\includegraphics[height=#1]{#3}}%
  {O}{l}{F}{F}{L}%
}
\newcommand\addunderlay[4]{%
  \stackengine{0pt}%
  {\kern#2\includegraphics[height=#1]{#4}}%
  {#3}%
  {O}{l}{F}{F}{L}%
}

\begin{document}
 
\title{Identifying Time Scales in Particle Production from Fields}

\author{Matthias Diez}
\affiliation{Institute of Physics, University of Graz, NAWI Graz, , 
Universitätsplatz 5, 8010 Graz, Austria}

\author{Reinhard Alkofer}
\affiliation{Institute of Physics, University of Graz, NAWI Graz, , 
Universitätsplatz 5, 8010 Graz, Austria}

\author{Christian Kohlf\"urst}
\affiliation{Helmholtz-Zentrum Dresden-Rossendorf, 
Bautzner Landstra{\ss}e 400, 01328 Dresden, Germany,}

\date{\today}

\begin{abstract}
Particle production through ultra-strong electric fields is a well-studied research field. 
Nevertheless, despite repeated attempts to relate the production rate within the field to the formation time of a particle, the latter is still shrouded in mystery.
We provide an interpretation of a particle distribution at finite times enabling us to isolate and, therefore, identify the 
relevant time scales regarding particle formation in quantum physics within and beyond perturbation theory. \\ [1mm]
\end{abstract}

\maketitle

\paragraph{Introduction.--} 

When subject to ultra-strong electric fields quantum electrodynamics (QED) predicts the vacuum to become a mediator for effective electromagnetic interactions. 
This leads to the emergence of non-linear couplings between electromagnetic fields. In turn, the vacuum itself becomes unstable creating the opportunity for electron-positron pairs to form \cite{HeisenbergEuler, Weisskopf}, see also the recent reviews on strong-field QED \cite{DiPiazza:2011tq,Gelis:2015kya,Fedotov}. 

Extreme field strengths for, at least, mesoscopically large spatial and temporal extensions 
are needed in order to observe real photons to interact in such a way. As the electron is the lightest electrically charged particle the corresponding energy scale is set by the electron mass $m_e$ and consequently the one for the electric field by the so-called critical electric field strength $E_{\rm cr.} = m_e^2c^3/e\hbar \approx 1.3 \times 10^{18}$ V/m with $e$ being the magnitude of the electron charge. 
It is challenging to achieve such field strengths in the laboratory, and thus experimental strong-field QED is still in its infancy. However, recent and future technological advancements in laser physics will eventually provide the opportunity to probe the ultrahigh-intensity sector of quantum electrodynamics (QED) in a controlled environment \cite{footnote}. 

An intricate matter in this regard is the issue of ``when'' a particle is formed. 
Within Quantum Field Theory (QFT) particle scattering is well understood in terms of the asymptotic in- and out-states. 
In this approach, however, fundamental questions including the one about the formation time of a particle remain unanswered. To this end we note that the latter question can be accurately defined for the case of pair production from fields \cite{Whitepaper, Thesis} but it is evident that this issue transcends beyond vacuum pair production \cite{PhysRevLett.122.023603}. 
Hence, any new insight might not only challenge our perception of time in scattering theory and, 
in particular, time evolution on the quantum level \cite{PhysRevLett.53.115}. 
It may very well reveal a new layer in quantum non-equilibrium physics and therefore may have an impact on 
all research areas that revolve around the ``birth'' of \mbox{(quasi-)particles}. Those areas include, in particular, astrophysics \cite{RUFFINI20101}, QED plasma physics \cite{PhysRevLett.101.200403, doi:10.1063/1.5144449, RevModPhys.94.045001}, quantum simulators \cite{PhysRevLett.111.201601} and solid state physics \cite{doi:10.1126/science.abi8627}. 
Last but not least, with respect to the field of relativistic heavy ion collisions it is desirable to establish and to understand the time scales of the Schwinger effect as this very likely will contribute to elucidate the pathway from non-equilibrium expansion to particle formation and eventually local equilibrium (hydrodynamic description) \cite{Huovinen:2006jp}. 

In this Letter we assess the concept of a 'particle formation time' building on a recent idea which challenged the notion that at intermediate times quantities in quantum field theory are unphysical and merely a means of calculation \cite{PhysRevD.105.016021}. 
Instead it was argued that quantities based on the adiabatic particle number could potentially reveal information with respect to quantum non-equilibrium physics, { {\it cf.\/} Ref.~\cite{Aleksandrov:2022wcz} for a discussion on experimental prospects}. For this investigation we applied the equal-time Wigner formalism (quantum kinetic theory) which provides access to non-perturbative particle formation at all times \cite{Hebenstreit, Kluger, Schmidt}. Together with the concept of the adiabatic particle number this 
method allows to identify time scales in pair production.

The outline is as follows: First we introduce the key elements of the equal-time Wigner formalism based on a simple-to-use toy model and provide context for the 'time evolution' in a quantum system. In this regard, we introduce an interpretation of quantum field theoretical results at intermediate times. On this evidence, we then characterize the different time scales that are relevant for pair production followed by a brief conclusion on the applicability of our observations.
%
We use throughout this letter units such that $\hbar=1$ and $c=1$.

\paragraph{Semi-classical approach.--} 

Experiments on electron-positron pair production using high-power lasers involve multiple time scales: 
The repetition rate of a laser is on the order of seconds, an ``ultra-short'' laser pulse refers to the femto- or attosecond
regime, and the QED time scale is given by the Compton time of an electron, 
$t_c = \hbar / m_e = 1.3 \times 10^{-21}$s=1.3 zs. Hence, any theoretical approach is forced to incorporate physics taking place on vastly different time scales.

One approach to resolve this issue is to put the laser scale in the focus and fully resolve particle and wave propagation through semi-classical methods \cite{PhysRevE.92.023305}. 
As the QED time scale is several orders of magnitude smaller quantum effects are regarded as instantaneous. 
In this context it is assumed that a local scattering event triggers the creation of an electron-positron pair which is always regarded as real and thus immediately subject to external forces \cite{PhysRevLett.67.2427}. This yields an efficient method to compute macroscopic observables. 

The 'cost' of such an approach is, of course, the complete loss of information on all small scales. Questions on when and how a particle is formed cannot be answered this way. Furthermore, even tiny deviations on the QED scale between theory and experiment are amplified heavily when embedded on the macroscopic scale. Resolving QED processes precisely is thus paramount.

\paragraph{Quantum aspect of pair production.--} 

To resolve the intricacies of particle creation we fully commit to an analysis resolving the Compton time scale. Then the non-Markovian nature of particle  formation becomes evident, as the entire history of a quantum system is crucial for the final state \cite{PhysRevD.78.061701}. Additionally, within background fields the boundary between quantum fluctuations and real (detectable) particles is blurred.

In terms of electron-positron pair production two fundamental building blocks are to be considered: the Schwinger effect \cite{Schwinger:1951nm, Sauter:1931} and multiphoton pair production \cite{BreitWheeler, Reiss}. The first mechanism requires ultra-strong electric fields. The underlying picture is that particles come into existence due to the background field doing work on virtual quantum fluctuations making it possible for them to cross the mass gap to become real particles. For multiphoton pair production, however, virtual pairs acquire energy through the high-frequency modes of a field. Again, a particle is considered 'real' if electron and positron overcome the threshold defined by their rest mass $2m_e$.

Functionally, the difference between these two mechanism is that the Schwinger effect is always exponentially suppressed,
number densities of produced pairs being $\propto \exp(-\pi m_e^2/|e|E)$,  with $E$ being the background field. 
Schwinger pair creation is driven by the low-frequency modes in the field which need to be of large amplitudes to obtain an effect. In contrast, multiphoton pair production has a perturbative sector such that production rates scale polynomially like $E^{2n}$ with $n$ being the number of scattered photons or, in the context of background fields, with the number of high-frequency modes of order $\mathcal O(m_e)$ \cite{Kohlfurst:2021skr}.



\paragraph{The model.--} 

Within the Wigner formalism we can isolate the particle formation process in such a way that no dynamical photons are taken into account. 
Moreover, the goal of this proof-of-concept study is to identify the relevant time scales in strong-field particle production and convey the information in the most transparent way. Hence, instead of more realistic scenarios we employ a field configuration,
\begin{equation}
|e| \mathbf{E} \left( t,x \right) = 
\varepsilon \ m_e^2 \ {\rm sech}^2 \left(t/\tau \right) \ 
\exp \left( -\frac{x^2}{2 \lambda^2}  \right) \mathbf{e}_x 
\, ,
 \label{eq:A}
\end{equation}
where quantum interference is minimal \cite{PhysRevLett.108.030401, PhysRevD.98.056009, PhysRevLett.101.130404} and the separation of charge carriers is easily recognized \cite{HebenstreitSelf}. Last but not least, this ansatz reduces the problem to 1+1 dimensions.
Herein, within the chosen units, the peak field strength $\varepsilon$ is a number given as a fraction of the 
critical field strength, $\tau$ parameterizes the pulse length, and $\lambda$ its spatial extension.
  
Within the equal-time Wigner formalism \cite{Vasak:1987umA, BB, Ochs} pair production in $1+1$ dimensions 
is described by \cite{PhysRevD.97.036026}
\begin{align}
  &D_t \mathbbm{s} - 2p \mathbbm{t}_{x} = 0,&& \ \label{eq1}
  D_t \mathbbm{v}_0 + \partial_x \mathbbm{v}_{x} = 0, \\
  &D_t \mathbbm{v}_x + \partial_x \mathbbm{v}_{x} = -2m_e \mathbbm{t}_{x}, && \
  D_t \mathbbm{t}_{x} + 2p \mathbbm{s} = 2m_e \mathbbm{v}_x, \label{eq2}
\end{align}
wherein the pseudo-differential operator on the left-hand side is 
$D_t = \partial_t + e \int_{-1/2}^{1/2} {\rm d} \xi E(t,x+{\rm i} \xi \partial_p) \partial_p$. Vacuum initial conditions are given by 
\begin{equation}
 \mathbbm{s}_{\rm vac} = -m_e/\sqrt{m_e^2+p^2} \, , \quad \mathbbm{v}_{x,\rm vac} = -p/\sqrt{m_e^2+p^2} \, ,
\end{equation}
and zero otherwise. The particle distribution function is given by
\begin{equation}
n \left( x, p \right) = \frac{m_e \left( \mathbbm{s}-\mathbbm{s}_{\rm vac} \right) + p \left(
\mathbbm{v}_x-\mathbbm{v}_{x, \rm vac} \right)}{\sqrt{m_e^2+p^2}}.
\end{equation}
Spectral and spatial distribution are obtained by integrating over the respective other coordinate. 
{The particle density $N$, which is obtained through integration over the whole phase-space, is thus derived from first principles \cite{Bloch, PhysRevD.83.025011,PhysRevD.79.065027}.}
%
We should make it absolutely clear though, that the concept of a time-evolution is frame-dependent \cite{PhysRevD.94.065005} and a 'particle number' that only counts observable particles is a purely classical concept, {\it cf.\/} Ref.~\cite{PhysRevD.83.025011}. Nevertheless, through careful assessment of numerical results we can still extract physically relevant information even at intermediate times.  


\paragraph{Time evolution.--} 
  
To study particle formation via the time evolution of the distribution function we first have to define what qualifies as a 'particle'. From the standard treatment of QFT the concepts of {\it virtual quantum fluctuations} and {\it real particles} \mbox{( = localized wave-packets build 
from asymptotic states)} are well-known. To simplify the description of structures in the particle distribution at intermediate times we introduce a third concept which we are calling {\it pre-particle}. We will argue that it is possible to give it a precise physical meaning even for fields with a still sizeable time dependence, and we are defining it by identifying that its motion coincides with the (semi-)classical expectations for the background field under
consideration.

\begin{figure}[t]
\includegraphics[width = 0.49\textwidth]{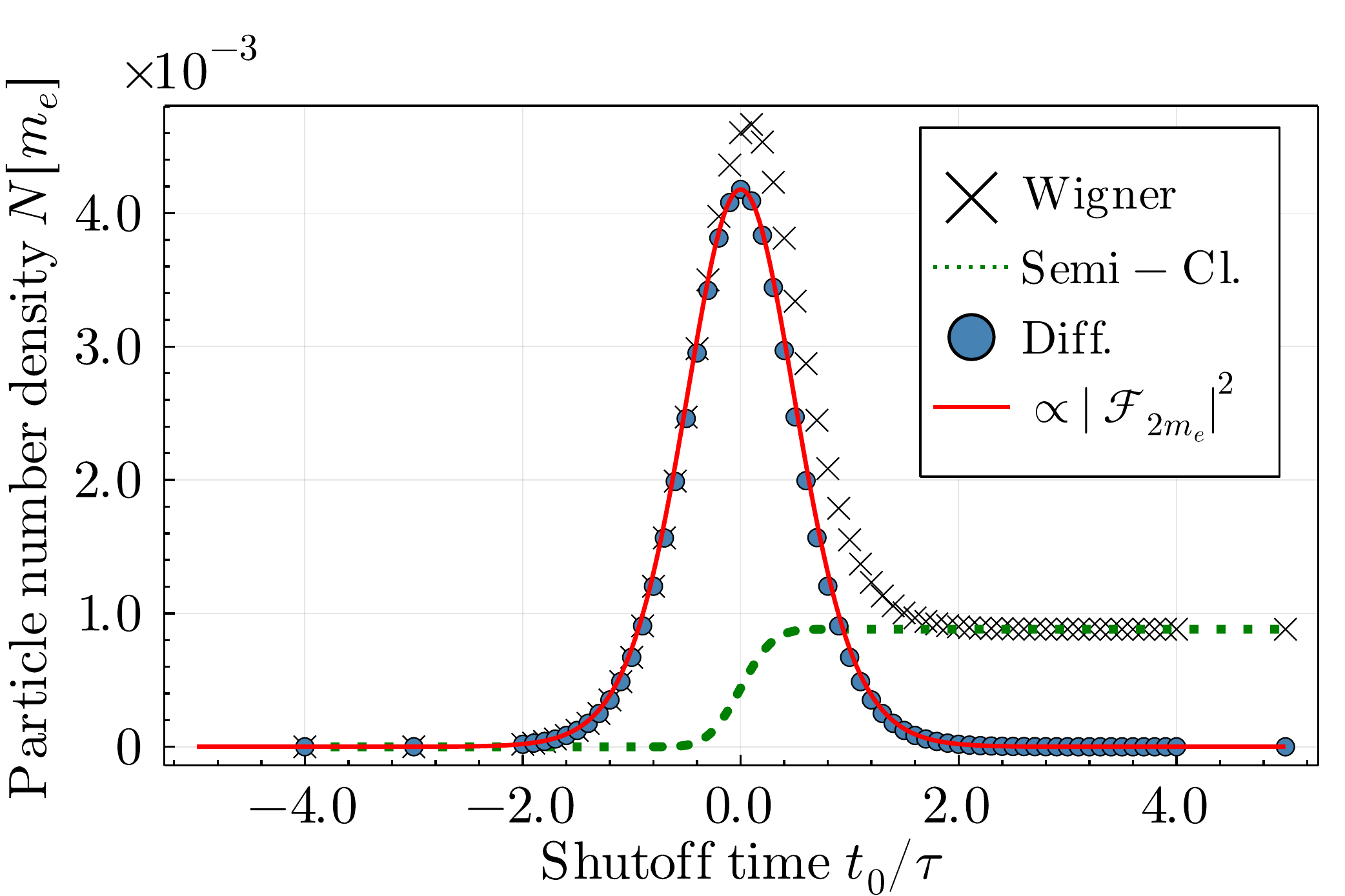}
\caption{ 
{Particle density $N$ as a function of the shutoff time $t_0$ normalized by the pulse length $\tau=30/m_e$ for a homogeneous Sauter pulse with peak strength $\varepsilon = 0.4$. Markers 'x' denote the particle density obtained through solving the Wigner transport equations, the green dotted curve corresponds to the semi-classical expectation values and the blue, filled circles display the difference of these solutions. For comparison, the red line represents the amplitude of the field's Fourier mode at frequency $2m_e$ times a scaling factor.}
}
\label{fig:ncreate}
\end{figure}

Semi-classical models for Schwinger pair production employ a very intuitive picture of particle creation: 
The force of the background electric field separates two point particles created by quantum tunneling through the mass gap \cite{PhysRevD.20.179}.
For the single-pulse toy model, Eq.~\eqref{eq:A}, this is a gradual process that calls first for a steep increase in the particle number (when the electric field becomes stronger) followed by a flattening of the curve (when the electric field becomes weaker again). Overall, at all points in time the particle number is expected to climb. This, however, is not how a quantum system evolves as is evident from Fig.~\ref{fig:ncreate}. 
For an exemplary set of parameters $\varepsilon$, $\tau$ (and $\lambda$)
the 'particle' number overshoots the expectation value by 
a factor of four  only to drop-off and eventually converge towards the semi-classical result at final times 
$t \gg \tau$. For smaller field strengths this behaviour is even more extreme as a disparity of several orders of magnitude can be achieved \cite{PhysRevD.83.025011}. 

\begin{figure}[t]
\includegraphics[width=0.49\textwidth]{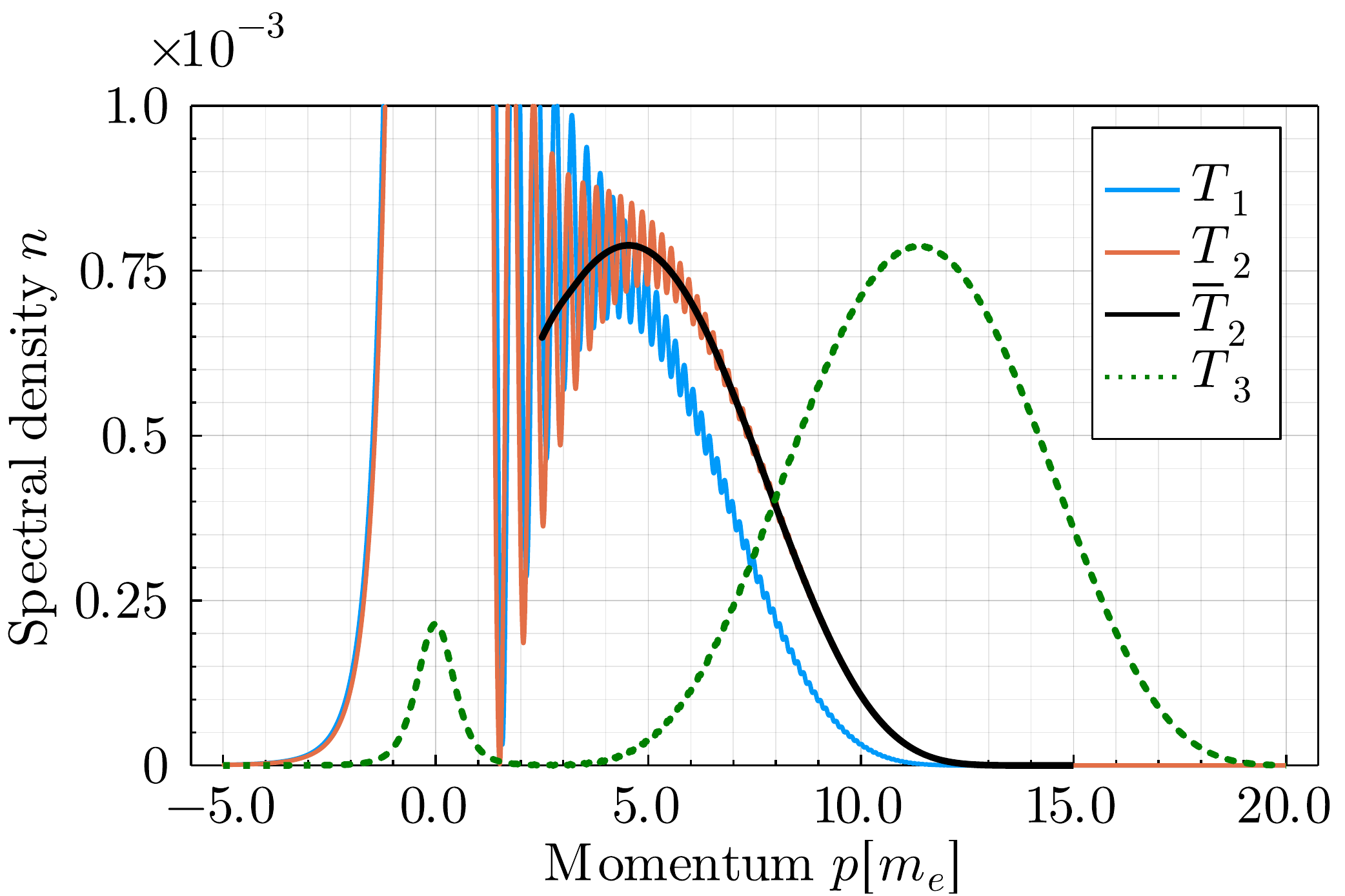}
\caption{
{Spectral particle distribution at times $T_1$ (build up) in blue, $T_2$ (separation) in red and $T_3$ (consolidation) in green for the homogeneous Sauter field with $\varepsilon=0.4$ and $\tau=30/m_e$. We display a moving average $\overline T_2$ (black line) to reveal the separation process amidst the quantum interference pattern. At times $T_3$ the peak at $p=0$ has become negligible having dropped to $1\%$ of its maximal value.} 
}
\label{fig:time-evolution}
\end{figure}

The fact that semi-classical expectations disagree with quantum field theoretical observations is further reinforced by the behaviour of the particle spectrum as a function of time, {\it cf.\/} Fig.~\ref{fig:time-evolution}. Instead of a single-peaked function to gradually form over time (and to be simultaneously accelerated by the field) we observe several structures in the momentum spectrum, typically not visible in coordinate space. 
First, a peak around vanishing momentum takes form (this coincides with the extreme increase in the yield in Fig.~\ref{fig:ncreate}). This peak is not accelerated by the electric field as one would expect for charged particles. Instead it lingers and then fades away again thereby seemingly giving rise to a second peak that emerges at non-zero momentum. At its first appearance this second peak is masked by oscillations reflecting a pattern of constructive and destructive interference. Nevertheless, the related underlying wave-packet is indeed subject to the Lorentz force and thus, even more important, moves out of the creation zone along the classical trajectory. It is therefore the second peak which can be identified with the pre-particle.
 
The solution to this seeming conundrum is to realize that at intermediate times a particle does neither 'exist' nor 'not exist' but due to the probabilistic nature of quanta its existence is conditional. Together with the fact that such QFT processes are non-Markovian this allows for a very specific interpretation: Suppose a strong field has built up and is present until a time $t_0$ then a pair might be created with 
probability $P_0$  but as the then future shape of the field is unknown the asymptotically late (final) creation probability $P^{\mathrm asymp}$ can be very different from $P_0$.
This is true except for one case, namely, if the field is shaped very similar to $E(t,x) \Theta(t_0-t)$. Suppose now that
up to $t_0$ the electric background field is given by the toy model Eq.~(\ref{eq:A}) and then switched off very rapidly, resp., shutoff. The pair creation probability can then be estimated quite precisely by $P_0$. As a matter of fact, the calculated spectrum will correctly represent the then measured spectrum. We observed, however, that if the field (\ref{eq:A}) is not switched off much less particles would have been produced. This is to be expected because due to the shutoff the field contains high-frequency modes which give rise to multiphoton pair production. 
Hence, instead of trying to interpret the results at each point in time as possibly detectable signals (which they are not due to the interplay of the background field's time dependence and the uncertainty principle), the time evolution of the quantum system \eqref{eq1}-\eqref{eq2} should be  considered as a sequence of 'What if?'-events. At each instant in time we may observe the spectrum that would have been created if the background field were to drop to zero practically instantaneously \cite{PhysRevD.105.016021}. 

Therefore, as the origin of the virtual fluctuation at $p \approx 0$ and the pre-particle are quite different we can verify the above picture by analyzing our numerical results. According to the discussion above shutoff-induced signals in the particle distribution stem from a virtual excitation created by modes with a frequency close or identical to the threshold for pair production, $\omega = 2m_e$. Therefore, the amplitude of the Fourier transform {${\mathcal F}_{2m_e}$} of the field Eq.~\eqref{eq:A} amended by a shutoff at time $t_0$, $E(t,x) \Theta(t_0-t)$, evaluated at $2m_e$ (leading order perturbative contribution)
directly corresponds to the fluctuating fraction in the particle signal. We verified this explicitly 
in our numerical results, see Fig.~\ref{fig:ncreate}, in which we display the relation between the particle number and perturbative contribution as a function of shutoff time $t_0$.


\paragraph{Particle lifetime, formation time and coherence time.--} 

We have established a routine to identify pre-particles that are destined to become real at asymptotic times in contrast to signals emerging due to a possible shutoff in the field or due to quantum fluctuations in general. On this basis we can provide a quantitative measurement of time scales for the quantum field theoretical system, Eqs.~\eqref{eq1}-\eqref{eq2}.

Two scales we can already obtain on the level of simple quantum mechanics, {\it cf.\/} a discussion in terms of a quantum Boltzmann approach \cite{PhysRevD.50.6911}. First, the Heisenberg uncertainty principle provides a first time scale corresponding to the 'lifetime' of a virtual electron-positron pair fluctuation, $\Delta \tau_{\rm lifetime} = \hbar/2m_e$. 
Second, relating the work done by the field over time with the threshold for particle creation the concept of a coherence time $\Delta \tau_{\rm coherence} = 1/m_e\varepsilon$ can be introduced \cite{doi:10.1007/BF01120220}. For example, at critical field strengths it takes the field only the Compton time $1/m_e$ to break the bond between a virtual electron-positron pair.

Our focus is on the formation time, the time it takes a particle, resp., a pre-particle to be recognizable as such. 
Fig.~\ref{fig:time-evolution} serves as a blueprint for the following discussion illustrating that the development of a pre-particle is a process characterized by several time scales. The first narrow peak, which appears at a time when the electric field reaches its maximum, arises due to the uncertainty in the frequency spectrum of the electric field (high-frequency shutoff modes). In Schwinger pair production this fluctuation will eventually fade away and can thus be ignored for the purpose of determining the formation time of the (pre-)particle. After a time interval $T_1 \approx 0.65 \, \tau^{1/4} \, t_c^{3/4}
/\varepsilon^2$ an interference pattern starts to build up resulting in the appearance of a peak centered around a non-vanishing momentum. This happens due to the superposition of two particle distributions of different origin (shutoff versus genuine Schwinger). 

At first this  interference becomes stronger over time because more and more pre-particles are created through the Schwinger mechanism. Additionally, the Lorentz force dictates the acceleration of \mbox{(pre-)}particles. In strong background fields in the shape of Eq.~\eqref{eq:A} (pre)-particles accelerate quickly thus creating a clearly discernible signal in the particle distribution. The corresponding time scale for this separation is found to be $T_2=T_1+0.06 {\tau^{3/4} t_c^{1/4}/\varepsilon^{3/2}}$. At this and later times the background field strongly decreases and a would-be shutoff were to produce less particles. Correspondingly, the peak at $p \approx 0$ slowly disappears on the order of $T_3 \approx 1.8\tau$ at which it has decreased to 1\% of its maximal value while the pre-particle signal consolidates.   
 
\begin{figure}[t]
\includegraphics[width=0.49\textwidth]{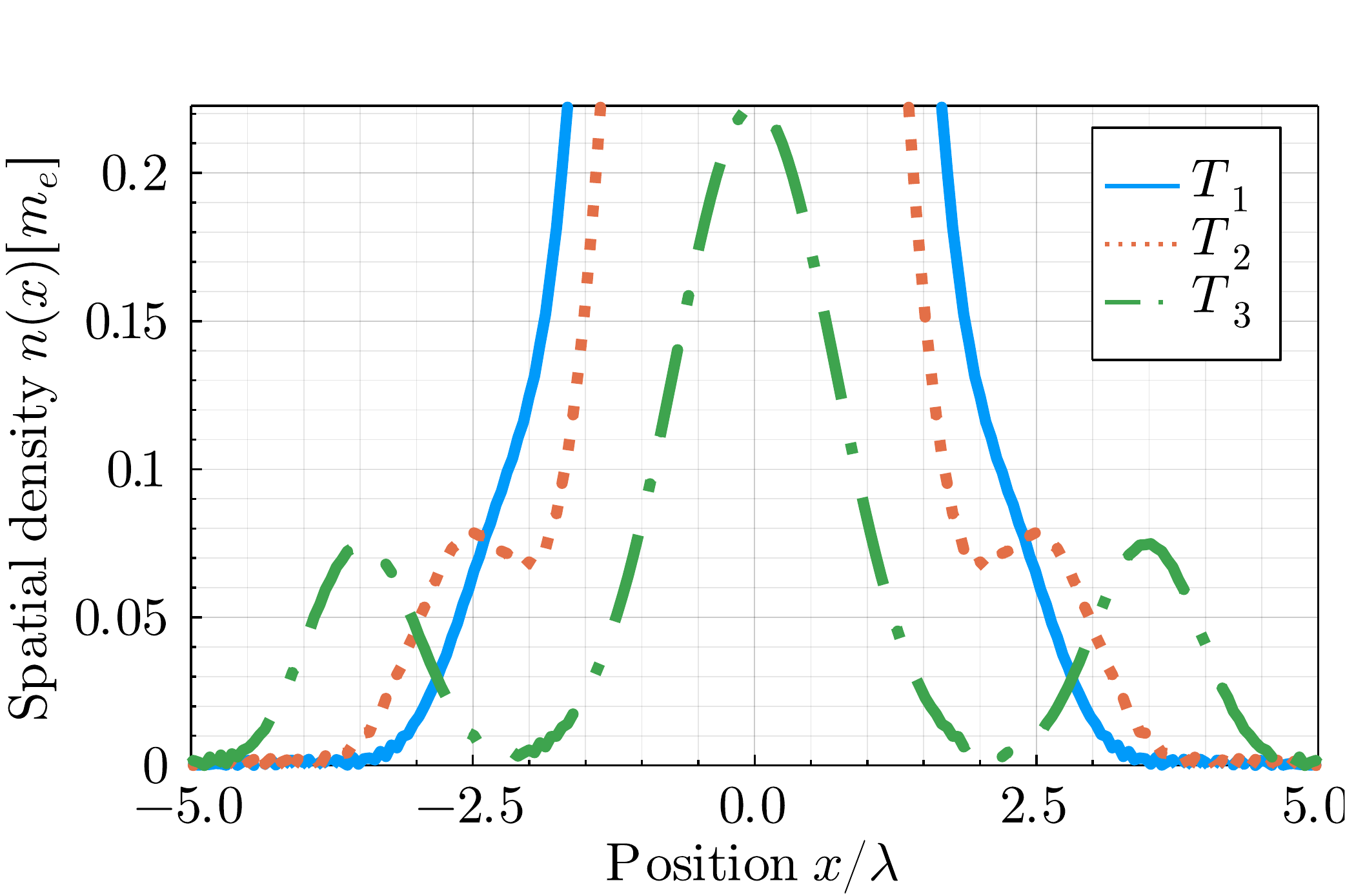}
\caption{
{Spatial density at  different times as a function of the normalized coordinate $x/\lambda$ for an inhomogeneous Sauter pulse with $\varepsilon=0.3$, $\tau=20/m_e$ and $\lambda=10/m_e$. For this specific set of parameters build up (blue) and separation (red) of the pre-particles at times $T_1$ and $T_2$, respectively, are observed at a distance $\sim 2.5\lambda$ from the origin. At $T_3$ the peak at $p \approx 0$ still has not fallen below the level of the pre-particles despite having retained only $1\%$ of its original size. Nevertheless, at times $T_3$ pre-particle and central peak form cleanly separated structures. 
}
}
\label{fig:inhom}
\end{figure}

Furthermore, a 'formation length' or 'particle distance at origin' can be attributed to the pair production process.
Such a classification establishes the related length scales, {\it cf.\/} its usage in terms of photon emission \cite{BAIER2005261, RevModPhys.77.1131}. Fig.~\ref{fig:inhom} serves as a guide for such an analysis. The peak formed around $p \approx 0$ is again the signature of a would-be shutoff high-frequency contribution that blocks the view on the pre-particle formation. Nevertheless, after a time $T_2$ the narrow peak stemming from a possible shutoff is accompanied by two additional side peaks. These correspond to the pre-particles that have formed due to the strong field facilitating Schwinger particle creation.  
Note that in coordinate space, the specific shape of the electric field \eqref{eq:A} helps us to easily distinguish pre-electrons from pre-positrons.

\paragraph{Conclusion.--} 
 
In this Letter we have examined the question whether a quantum distribution function can hold sensible physical information even at finite, pre-asymptotic times. According to the presented interpretation the distribution function does indeed relate to a physical, 'real' particle content. However, it also incorporates information that correlate with the particle distribution a suddenly switched-off field would generate. The latter vanishes asymptotically, but at intermediate times the superposition of these two vastly different parts of the amplitude create further quantum interference. 
Nevertheless, a pre-particle following in its motion the classical trajectory can be identified at quite early times. \cite{footnote2} 

Building on this idea we were able to derive three relevant time scales for Schwinger pair production: \\
1) time scale $T_1 \approx 0.65 \, \tau^{1/4} \, t_c^{3/4} /\varepsilon^2$ for the appearance of a peak in the particle distribution related to a pre-particle; 
2) time scale $T_2=T_1+0.06 \tau^{3/4} t_c^{1/4}/\varepsilon^{3/2}$ at which the pre-particle as such starts to follow the classical trajectory; and
3) time scale $T_3=1.8 \tau$ at which quantum fluctuations (quantum interference and the `perturbative` peak related to a would-be shutoff) fade out. 

\paragraph{Acknowledgements.--} 

We are grateful to the organizers of the MP3 Multi-Petawatt Physics Prioritization Workshop for bringing the community together for interesting discussions on the future of strong-field quantum electrodynamics. We thank Ralf Sch\"utzhold and Holger Gies for helpful comments on the manuscript.




\begin{thebibliography}{99}
 \bibitem{HeisenbergEuler} 
 W.~Heisenberg and H.~Euler,
 {\em Consequences of Dirac's theory of positrons},
 Z.\ Phys.\ {\bf 98}, 714 (1936). 
 %
 \bibitem{Weisskopf} 
 V.~Weisskopf,
 {\em The electrodynamics of the vacuum based on the quantum theory of the electron}, 
 Kong.\ Dan.\ Vid.\ Sel.\ Mat.\ Fys.\ Med.\ {\bf 14N6}, 1 (1936).
\bibitem{DiPiazza:2011tq}
A.~Di Piazza, C.~Muller, K.~Z.~Hatsagortsyan and C.~H.~Keitel,
{\em Extremely high-intensity laser interactions with fundamental quantum systems},
Rev.\ Mod.\ Phys.\ \textbf{84}, 1177 (2012).
\bibitem{Gelis:2015kya}
F.~Gelis and N.~Tanji,
{\em Schwinger mechanism revisited},
Prog.\ Part.\ Nucl.\ Phys.\ \textbf{87}, 1 (2016).
\bibitem{Fedotov}
 A.~Fedotov, A.~Ilderton, F.~Karbstein, B.~King, D.~Seipt, H.~Taya and G.~Torgrimsson,
 {\em Advances in QED with intense background fields},
 arXiv:2203.00019 [hep-ph].
 \bibitem{footnote}
 Laser facilities with the capability to probe the strong-field QED sector are, e.g., HIBEF, LUXE, ELI, CILEX, CoReLS, ShenGuang-II as well as SLAC; 
 see the websites 
 
 \url{https://www.hibef.eu},  
 
 \url{https://luxe.desy.de}
 
 \url{https://www.eli-beams.eu},   
 
 \url{https://eli-laser.eu},
 
 \url{https://cilexsaclay.fr},
 
 \url{https://corels.ibs.re.kr},
 
 \url{https://lssf.cas.cn/en},
 
 \url{https://www.slac.stanford.edu}.
 \bibitem{Whitepaper}
 R.~Alkofer, M.~Diez, and C.~Kohlfürst,
 {\em What are the time-scales of particle formation in the Schwinger effect?},
 whitepaper for MP3 Multi-Petawatt Physics Prioritization Workshop, (2021),
 
 \url{https://mp3.lle.rochester.edu}.
 \bibitem{Thesis}
 M.~Diez,
 {\em Time-scales of particle formation in the Sauter-Schwinger effect},
 Master thesis, (2022),
 
 \url{https://unipub.uni-graz.at}.
 \bibitem{PhysRevLett.122.023603}
 Q.~Su and R.~Grobe,
 {\em Dirac Vacuum as a Transport Medium for Information},
 Phys.\ Rev.\ Lett.\ {\bf 122}, 023603 (2019).
 \bibitem{PhysRevLett.53.115}
 E.~Pollak and W.H.~Miller,
 {\em New Physical Interpretation for Time in Scattering Theory},
 Phys.\ Rev.\ Lett.\ {\bf 53}, 115 (1984).
 \bibitem{RUFFINI20101}
 R.~Ruffini, G.~Vereshchagin, and S.-S.~Xue,
 {\em Electron–positron pairs in physics and astrophysics: From heavy nuclei to black holes},
 Phys.\ Rep.\ {\bf 487}, 1 (2010). 
 \bibitem{PhysRevLett.101.200403} 
 A.R.~Bell and J.G.~Kirk,
 {\em Possibility of Prolific Pair Production with High-Power Lasers},
 Phys.\ Rev.\ Lett.\ {\bf 101}, 200403 (2008).
 %
 \bibitem{doi:10.1063/1.5144449}
 P.~Zhang, S.S.~Bulanov, D.~Seipt, A.V.~Arefiev, and A.G.R.~Thomas,
 {\em Relativistic plasma physics in supercritical fields},
 Physics\ of\ Plasmas\ {\bf 27}, 050601 (2020).
 %
 \bibitem{RevModPhys.94.045001}
 A.~Gonoskov, T.G.~Blackburn, M.~Marklund, and S.S.~Bulanov, 
 {\em Charged particle motion and radiation in strong electromagnetic fields},
 Rev.\ Mod.\ Phys.\ {\bf 94}, 045001 (2022). 
 \bibitem{PhysRevLett.111.201601}
 F. Hebenstreit, J. Berges, and D. Gelfand,
 {\em Real-Time Dynamics of String Breaking},
 Phys.\ Rev.\ Lett.\ {\bf 111}, 201601 (2013).
 \bibitem{doi:10.1126/science.abi8627}
 A.I.~Berdyugin, N.~Xin, H.~Gao, S.~Slizovskiy, Z.~Dong, S.~Bhattacharjee, P.~Kumaravadivel, S.~Xu, L.A.~Ponomarenko, M.~Holwill, D.A.~Bandurin, M.~Kim, Y.~Cao, M.T.~Greenaway, K.S.~Novoselov, I.V.~Grigorieva, K.~Watanabe, T.~Taniguchi, V.I.~Fal’ko, L.S.~Levitov, R.K.~Kumar, A.K.~Geim,
 {\em Out-of-equilibrium criticalities in graphene superlattices},
 Science\ {\bf 375}, 6579 (2022).   
 \bibitem{Huovinen:2006jp}
 P.~Huovinen and P.V.~Ruuskanen,
 {\em Hydrodynamic Models for Heavy Ion Collisions},
 Ann.\ Rev.\ Nucl.\ Part.\ Sci.\ {\bf 56}, 163 (2006).
 \bibitem{PhysRevD.105.016021}
 A.~Ilderton,
 {\em Physics of adiabatic particle number in the Schwinger effect},
 Phys.\ Rev.\ D\ {\bf 105}, 016021 (2022).
 \bibitem{Aleksandrov:2022wcz}
 I.A.~Aleksandrov, D.G.~Sevostyanov and V.M.~Shabaev,
 {\em Schwinger particle production: rapid switch off of the external field versus dynamical assistance},
 arXiv:2210.15626 [hep-ph].
 \bibitem{Hebenstreit}
 F.~Hebenstreit, R.~Alkofer and H.~Gies,
 {\em Schwinger pair production in space- and time-dependent electric fields: Relating the Wigner formalism to quantum kinetic theory},
 Phys.\ Rev.\ D {\bf 82}, 105026 (2010).
 %
 \bibitem{Kluger}
 Y.~Kluger, J.M.~Eisenberg, B.~Svetitsky, F.~Cooper and E.~Mottola,
 {\em Fermion pair production in a strong electric field},
 Phys.\ Rev.\ D {\bf 45}, 4659 (1992).
 %
 \bibitem{Schmidt} 
 S.M.~Schmidt, D.B.~Blaschke, G.~R\"opke, S.A.~Smolyansky, A.V.~Prozorkevich and V.D.~Toneev,
 {\em A quantum kinetic equation for particle production in the Schwinger mechanism},
 Int.\ J.\ Mod.\ Phys.\ E {\bf 7}, 709 (1998).  
 \bibitem{PhysRevE.92.023305}
 A.~Gonoskov, S.~Bastrakov, E.~Efimenko, A.~Ilderton, M.~Marklund, I.~Meyerov, A.~Muraviev, A.~Sergeev, I.~Surmin, and E.~Wallin
 {\em Extended particle-in-cell schemes for physics in ultrastrong laser fields: Review and developments},
 Phys.\ Rev.\ E\ {\bf 92}, 023305 (2015).
 \bibitem{PhysRevLett.67.2427}
 Y.~Kluger, J.M.~Eisenberg, B.~Svetitsky, F.~Cooper, and E.~Mottola,
 {\em Pair production in a strong electric field},
 Phys.\ Rev.\ Lett.\ {\bf 67}, 2427 (1991).
 \bibitem{PhysRevD.78.061701}
 F.~Hebenstreit, R.~Alkofer, and H.~Gies,
 {\em Pair production beyond the Schwinger formula in time-dependent electric fields},
 Phys.\ Rev.\ D\ {\bf 78}, 061701(R) (2008).
 \bibitem{Schwinger:1951nm} 
 J.~S.~Schwinger,
 {\em On gauge invariance and vacuum polarization},
 Phys.\ Rev.\ {\bf 82}, 664 (1951).
 %
 \bibitem{Sauter:1931} 
 F.~Sauter,
 {\em \"Uber das Verhalten eines Elektrons im homogenen elektrischen Feld nach der  relativistischen Theorie Diracs},
 Z.\ Phys.\ {\bf 69}, 742 (1931).
 \bibitem{BreitWheeler}
 G.~Breit and J.A.~Wheeler, 
 {\em Collision of Two Light Quanta},
 Phys.\ Rev.\ {\bf 46}, 1087 (1934).
 %
 \bibitem{Reiss}
 H.R.~Reiss, 
 {\em  Absorption of Light by Light},
 J.\ Math.\ Phys.\ {\bf 3}, 59 (1962).
 \bibitem{Kohlfurst:2021skr}
 C.~Kohlf\"urst, N.~Ahmadiniaz, J.~Oertel, and R.~Sch\"utzhold, 
 {\em Sauter-Schwinger effect for colliding laser pulses},
 arXiv:2107.08741 [hep-ph].
 \bibitem{PhysRevLett.108.030401}
 E.~Akkermans and G.V.~Dunne,
 {\em Ramsey Fringes and Time-Domain Multiple-Slit Interference from Vacuum},
 Phys.\ Rev.\ Lett.\ {\bf 108}, 030401 (2012).
 %
 \bibitem{PhysRevD.98.056009}
 J.Z.~Kamiński, M.~Twardy and K.~Krajewska,
 {\em Diffraction at a time grating in electron-positron pair creation from the vacuum},
 Phys.\ Rev.\ D {\bf 98}, 056009 (2018).
 %
 \bibitem{PhysRevLett.101.130404}
 R.~Schützhold, H.~Gies, and G.~Dunne,
 {\em Dynamically Assisted Schwinger Mechanism},
 Phys.\ Rev.\ Lett.\ {\bf 101}, 130404 (2008).
\bibitem{HebenstreitSelf}
 F.~Hebenstreit, R.~Alkofer and H.~Gies,
 {\em Particle Self-Bunching in the Schwinger Effect in Spacetime-Dependent Electric Fields},
 Phys.\ Rev.\ Lett.\ {\bf 107}, 180403 (2011). 
 \bibitem{Vasak:1987umA} 
 D.~Vasak, M.~Gyulassy and H.T.~Elze, 
 {\em Quantum Transport Theory for Abelian Plasmas},
 Ann.\ Phys.\ (N.Y.) {\bf 173}, 462 (1987).
 %
 \bibitem{BB} 
 I.~Bialynicki-Birula, P.~G\'ornicki and J.~Rafelski, 
 {\em Phase-space structure of the Dirac vacuum},
 Phys.\ Rev.\ D {\bf 44}, 1825 (1991).
 %
 \bibitem{Ochs}
 S.~Ochs and U.~Heinz,
 {\em Wigner Functions in Covariant and Single-Time Formulations},
 Ann.\ Phys.\ {\bf 266}, 351 (1998).
 \bibitem{PhysRevD.97.036026}
 C.~Kohlfürst and R.~Alkofer,
 {\em Ponderomotive effects in multiphoton pair production},
 Phys.\ Rev.\ D\ {\bf 97}, 036026 (2018). 
 \bibitem{Bloch} 
 J.C.R.~Bloch, V.A.~Mizerny, A.V.~Prozorkevich, C.D.~Roberts, S.M.~Schmidt, S.A.~Smolyansky and D.V.~Vinnik, 
 {\em Pair creation: Back reactions and damping},
 Phys.\ Rev.\ D {\bf 60}, 116011 (1999).  
 \bibitem{PhysRevD.83.025011}
 A.M.~Fedotov, E.G.~Gelfer, K.Yu.~Korolev, and S.A.~Smolyansky,
 {\em Kinetic equation approach to pair production by a time-dependent electric field},
 Phys.\ Rev.\ D\ {\bf 83}, 025011 (2011).
 %
 \bibitem{PhysRevD.79.065027}
 C.K.~Dumlu,
 {\em Quantum kinetic approach and the scattering approach to vacuum pair production},
 Phys.\ Rev.\ D\ {\bf 79}, 065027 (2009).
 \bibitem{PhysRevD.94.065005}
 R.~Dabrowski and G.V.~Dunne,
 {\em Time dependence of adiabatic particle number},
 Phys.\ Rev.\ D\ {\bf 94}, 065005 (2016).
 \bibitem{PhysRevD.20.179}
 A.~Casher, H.~Neuberger, and S.~Nussinov,
 {\em Chromoelectric-flux-tube model of particle production},
 Phys.\ Rev.\ D\ {\bf 20}, 179 (1979).
 \bibitem{PhysRevD.50.6911}
 J.~Rau,
 {\em Pair production in the quantum Boltzmann equation},
 Phys.\ Rev.\ D\ {\bf 50}, 6911 (1994).
 \bibitem{doi:10.1007/BF01120220}
 V.I.~Ritus,
 {\em Quantum effects of the interaction of elementary particles with an intense electromagnetic field},
 J.\ Russ.\ Laser\ Res.\ {\bf 6}, 497 (1985).
 \bibitem{BAIER2005261}
 V.N.~Baier and V.M.~Katkov,
 {\em Concept of formation length in radiation theory},
 Phys.\ Rept.\ {\bf 409}, 261 (2005).
 %
 \bibitem{RevModPhys.77.1131}
 U.I.~Uggerhøj,
 {\em The interaction of relativistic particles with strong crystalline fields},
 Rev.\ Mod.\ Phys.\ {\bf 77}, 1131 (2005).
 \bibitem{footnote2}
 Our analysis is applicable in the Schwinger regime $\tau \geq 10/m_e$, $0.1 \leq \varepsilon \leq 0.5$ and $\lambda \geq 10/m_e$. More details can be found in Ref. \cite{Thesis}. Further information will be provided in a future publication. 
\end{thebibliography}
\end{document}